\pdfoutput=1
\documentclass[12pt,a4paper]{article}
\usepackage{ifthen}
\usepackage{rotating}
\usepackage{lscape}

\newboolean{pdflatex}
\setboolean{pdflatex}{true}

\newboolean{articletitles}
\setboolean{articletitles}{true}

\newboolean{uprightparticles}
\setboolean{uprightparticles}{false}

\newcommand{\bck}{\Bcp\to\jpsi\Kp}
\newcommand{\bcp}{\Bcp\to\jpsi\pip}

\newcommand{\tm}{\ensuremath{\rm \,Tm}\xspace}

\usepackage{microtype}
\usepackage{lineno}  
\usepackage{xspace}

\usepackage{graphicx} 
\usepackage{color}
\usepackage{colortbl}
\graphicspath{{./figs/}}

\usepackage{amsmath} 
\usepackage{amssymb}
\usepackage{amsfonts}
\usepackage{upgreek} 

\newcommand*\patchAmsMathEnvironmentForLineno[1]{
\expandafter\let\csname old#1\expandafter\endcsname\csname #1\endcsname
\expandafter\let\csname oldend#1\expandafter\endcsname\csname
end#1\endcsname
 \renewenvironment{#1}
   {\linenomath\csname old#1\endcsname}
   {\csname oldend#1\endcsname\endlinenomath}
}
\newcommand*\patchBothAmsMathEnvironmentsForLineno[1]{
  \patchAmsMathEnvironmentForLineno{#1}
  \patchAmsMathEnvironmentForLineno{#1*}
}
\AtBeginDocument{
\patchBothAmsMathEnvironmentsForLineno{equation}
\patchBothAmsMathEnvironmentsForLineno{align}
\patchBothAmsMathEnvironmentsForLineno{flalign}
\patchBothAmsMathEnvironmentsForLineno{alignat}
\patchBothAmsMathEnvironmentsForLineno{gather}
\patchBothAmsMathEnvironmentsForLineno{multline}
}

\usepackage{hyperref}   
\usepackage[all]{hypcap} 

\def\lhcb {\mbox{LHCb}\xspace}
\def\ux85 {\mbox{UX85}\xspace}

\def\tevatron {Tevatron\xspace}

\def\rich   {RICH\xspace}

\ifthenelse{\boolean{uprightparticles}}
{

 \def\Ppi         {\ensuremath{\uppi}\xspace}

 \def\Ppsi        {\ensuremath{\uppsi}\xspace}

 \def\PDelta      {\ensuremath{\Delta}\xspace}
 \def\PXi      {\ensuremath{\Xi}\xspace}
 \def\PLambda      {\ensuremath{\Lambda}\xspace}
 \def\PSigma      {\ensuremath{\Sigma}\xspace}
 \def\POmega      {\ensuremath{\Omega}\xspace}
 \def\PUpsilon      {\ensuremath{\Upsilon}\xspace}

 \def\PB      {\ensuremath{\mathrm{B}}\xspace}
 
 \def\PD      {\ensuremath{\mathrm{D}}\xspace}

 \def\PJ      {\ensuremath{\mathrm{J}}\xspace}
 \def\PK      {\ensuremath{\mathrm{K}}\xspace}

 \def\Pc      {\ensuremath{\mathrm{c}}\xspace}

 \def\Pi      {\ensuremath{\mathrm{i}}\xspace}

 \def\Ps      {\ensuremath{\mathrm{s}}\xspace}

}
{

 \def\Ppi         {\ensuremath{\pi}\xspace}

 \def\Ppsi        {\ensuremath{\psi}\xspace}
 
 \mathchardef\PDelta="7101
 \mathchardef\PXi="7104
 \mathchardef\PLambda="7103
 \mathchardef\PSigma="7106
 \mathchardef\POmega="710A
 \mathchardef\PUpsilon="7107
 
 \def\PB      {\ensuremath{B}\xspace}
 
 \def\PD      {\ensuremath{D}\xspace}

 \def\PJ      {\ensuremath{J}\xspace}
 \def\PK      {\ensuremath{K}\xspace}

 \def\Pc      {\ensuremath{c}\xspace}

 \def\Pi      {\ensuremath{i}\xspace}

 \def\Ps      {\ensuremath{s}\xspace}

}

\def\squark    {\ensuremath{\Ps}\xspace}

\def\cquark    {\ensuremath{\Pc}\xspace}

\def\pion  {\ensuremath{\Ppi}\xspace}

\def\pip   {\ensuremath{\pion^+}\xspace}
\def\pim   {\ensuremath{\pion^-}\xspace}

\def\kaon  {\ensuremath{\PK}\xspace}

  \def\Kbar  {\kern 0.2em\overline{\kern -0.2em \PK}{}\xspace}

\def\Kz    {\ensuremath{\kaon^0}\xspace}
\def\Kzb   {\ensuremath{\Kbar^0}\xspace}
\def\KzKzb {\ensuremath{\Kz \kern -0.16em \Kzb}\xspace}
\def\Kp    {\ensuremath{\kaon^+}\xspace}
\def\Km    {\ensuremath{\kaon^-}\xspace}

\def\KpKm  {\ensuremath{\Kp \kern -0.16em \Km}\xspace}

  \def\Dbar    {\kern 0.2em\overline{\kern -0.2em \PD}{}\xspace}
\def\D       {\ensuremath{\PD}\xspace}

\def\Dz      {\ensuremath{\D^0}\xspace}
\def\Dzb     {\ensuremath{\Dbar^0}\xspace}
\def\DzDzb   {\ensuremath{\Dz {\kern -0.16em \Dzb}}\xspace}
\def\Dp      {\ensuremath{\D^+}\xspace}
\def\Dm      {\ensuremath{\D^-}\xspace}

\def\DpDm    {\ensuremath{\Dp {\kern -0.16em \Dm}}\xspace}

\def\Dsm     {\ensuremath{\D^-_\squark}\xspace}

\def\B       {\ensuremath{\PB}\xspace}

\def\Bbar    {\ensuremath{\kern 0.18em\overline{\kern -0.18em \PB}{}}\xspace}

\def\Bu      {\ensuremath{\B^+}\xspace}

\def\Bd      {\ensuremath{\B^0}\xspace}
\def\Bs      {\ensuremath{\B^0_\squark}\xspace}

\def\Bcp     {\ensuremath{\B_\cquark^+}\xspace}

\def\jpsi     {\ensuremath{{\PJ\mskip -3mu/\mskip -2mu\Ppsi\mskip 2mu}}\xspace}
\def\psitwos  {\ensuremath{\Ppsi{(2S)}}\xspace}

  \def\Y#1S{\ensuremath{\PUpsilon{(#1S)}}\xspace}

\def\Lbar {\ensuremath{\kern 0.1em\overline{\kern -0.1em\PLambda}}\xspace}

\def\BF         {{\ensuremath{\cal B}\xspace}}

\def\BR         {\BF}

\def\to                 {\ensuremath{\rightarrow}\xspace}

\def\AT#1     {\ensuremath{A_{\mathrm{T}}^{#1}}\xspace}           

\def\C#1      {\ensuremath{\mathcal{C}_{#1}}\xspace}                       
\def\Cp#1     {\ensuremath{\mathcal{C}_{#1}^{'}}\xspace}                    
\def\Ceff#1   {\ensuremath{\mathcal{C}_{#1}^{\mathrm{(eff)}}}\xspace}        
\def\Cpeff#1  {\ensuremath{\mathcal{C}_{#1}^{'\mathrm{(eff)}}}\xspace}       
\def\Ope#1    {\ensuremath{\mathcal{O}_{#1}}\xspace}             
\def\Opep#1   {\ensuremath{\mathcal{O}_{#1}^{'}}\xspace}

\newcommand{\tev}{\ensuremath{\mathrm{\,Te\kern -0.1em V}}\xspace}
\newcommand{\gev}{\ensuremath{\mathrm{\,Ge\kern -0.1em V}}\xspace}
\newcommand{\mev}{\ensuremath{\mathrm{\,Me\kern -0.1em V}}\xspace}
\newcommand{\kev}{\ensuremath{\mathrm{\,ke\kern -0.1em V}}\xspace}
\newcommand{\ev}{\ensuremath{\mathrm{\,e\kern -0.1em V}}\xspace}
\newcommand{\gevc}{\ensuremath{{\mathrm{\,Ge\kern -0.1em V\!/}c}}\xspace}
\newcommand{\mevc}{\ensuremath{{\mathrm{\,Me\kern -0.1em V\!/}c}}\xspace}
\newcommand{\gevcc}{\ensuremath{{\mathrm{\,Ge\kern -0.1em V\!/}c^2}}\xspace}
\newcommand{\gevgevcccc}{\ensuremath{{\mathrm{\,Ge\kern -0.1em V^2\!/}c^4}}\xspace}
\newcommand{\mevcc}{\ensuremath{{\mathrm{\,Me\kern -0.1em V\!/}c^2}}\xspace}

\def\mum  {\ensuremath{\,\upmu\rm m}\xspace}

\def\mub{\ensuremath{\rm \,\upmu b}\xspace}

\def\invfb   {\ensuremath{\mbox{\,fb}^{-1}}\xspace}

\def\mhz  {\ensuremath{{\rm \,MHz}}\xspace}
\def\khz  {\ensuremath{{\rm \,kHz}}\xspace}

\newcommand{\stat}{\ensuremath{\mathrm{\,(stat)}}\xspace}

\newcommand{\chisq}{\ensuremath{\chi^2}\xspace}

\newcommand{\chisqip}{\ensuremath{\chi^2_{\rm IP}}\xspace}

\newcommand{\chisqvtx}{\ensuremath{\chi^2_{\rm vtx}}\xspace}

\def\gsim{{~\raise.15em\hbox{$>$}\kern-.85em
          \lower.35em\hbox{$\sim$}~}\xspace}
\def\lsim{{~\raise.15em\hbox{$<$}\kern-.85em
          \lower.35em\hbox{$\sim$}~}\xspace}

\def\pt         {\mbox{$p_{\rm T}$}\xspace}

\def\dllkpi     {\ensuremath{\mathrm{DLL}_{\kaon\pion}}\xspace}

\def\evtgen     {\mbox{\textsc{EvtGen}}\xspace}
\def\pythia     {\mbox{\textsc{Pythia}}\xspace}
\def\bcvegpy    {\mbox{\textsc{Bcvegpy}}\xspace}

\def\geant      {\mbox{\textsc{Geant4}}\xspace}

\def\photos     {\mbox{\textsc{Photos}}\xspace}

\def\tell1  {TELL1\xspace}
\def\ukl1   {UKL1\xspace}

\usepackage{cite} 
\usepackage{mciteplus}

\begin{document}

\renewcommand{\thefootnote}{\fnsymbol{footnote}}
\setcounter{footnote}{1}

\begin{titlepage}
\pagenumbering{roman}

\vspace*{-1.5cm}
\centerline{\large EUROPEAN ORGANIZATION FOR NUCLEAR RESEARCH (CERN)}
\vspace*{1.5cm}
\hspace*{-0.5cm}
\begin{tabular*}{\linewidth}{lc@{\extracolsep{\fill}}r}
\ifthenelse{\boolean{pdflatex}}
{\vspace*{-2.7cm}\mbox{\!\!\!\includegraphics[width=.14\textwidth]{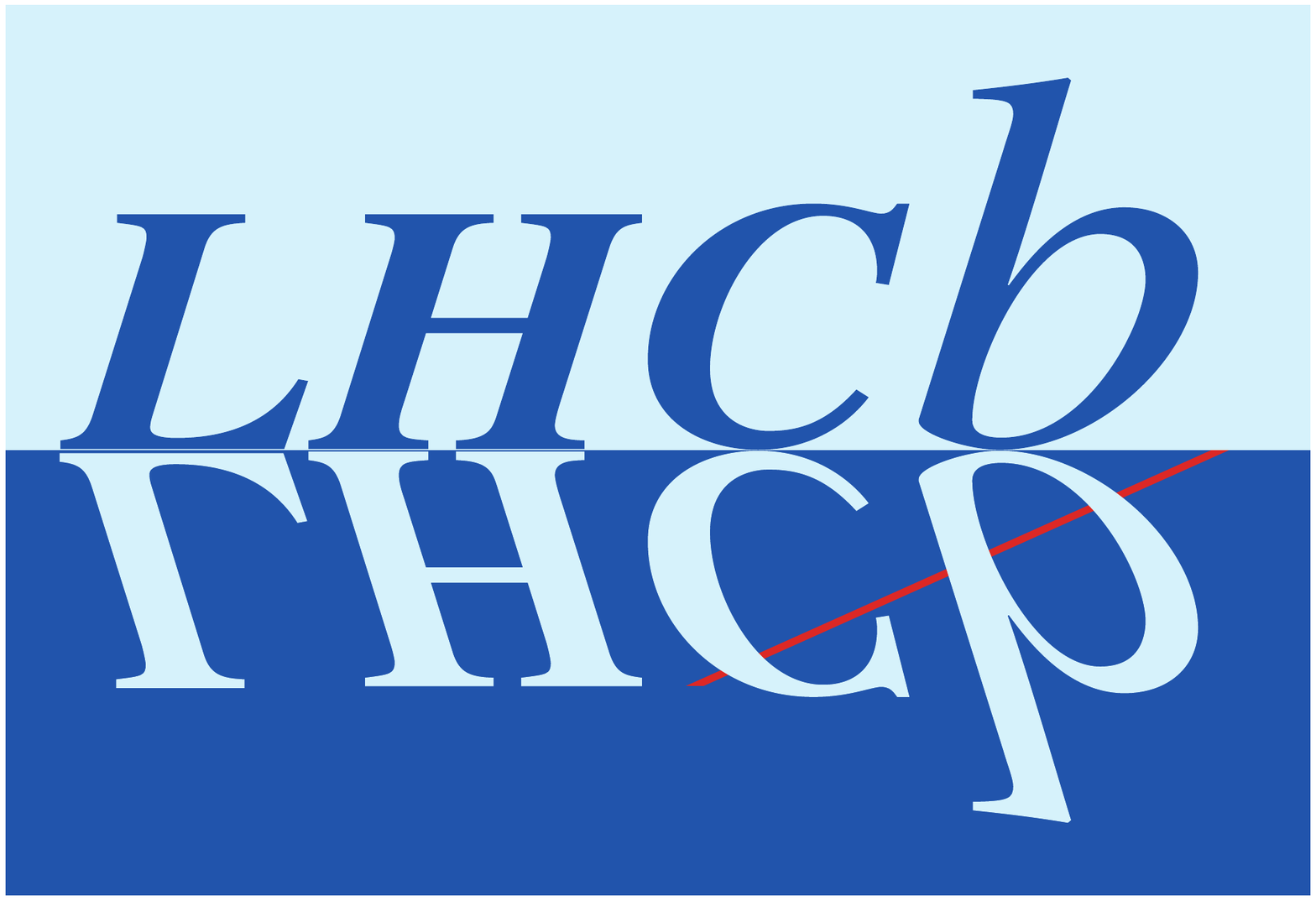}} & &}
{\vspace*{-1.2cm}\mbox{\!\!\!\includegraphics[width=.12\textwidth]{lhcb-logo.eps}} & &}
\\
 & & CERN-PH-EP-2013-106 \\
 & & LHCb-PAPER-2013-021 \\
 & & June 26 2013 \\
 & & \\

\end{tabular*}

\vspace*{4.0cm}

{\bf\boldmath\huge
\begin{center}
  First observation of the decay $\bck$
\end{center}
}

\vspace*{2.0cm}

\begin{center}
The LHCb collaboration\footnote{Authors are listed on the following pages.}
\end{center}

\vspace{\fill}

\begin{abstract}
  \noindent
  The decay $\bck$ is observed for the first time using a data sample, corresponding to an integrated luminosity of 1.0 \invfb, collected by the \lhcb experiment in $pp$ collisions at a centre-of-mass energy of 7\tev. A yield of $46\pm12$ events is reported, with a significance of 5.0 standard deviations. The ratio of the branching fraction of $\bck$ to that of $\bcp$ is measured to be $0.069\pm0.019\pm0.005$, where the first uncertainty is statistical and the second is systematic.

\end{abstract}

\vspace*{2.0cm}

\begin{center}
  Submitted to JHEP
\end{center}

\vspace{\fill}

{\footnotesize
\centerline{\copyright~CERN on behalf of the \lhcb collaboration, license \href{http://creativecommons.org/licenses/by/3.0/}{CC-BY-3.0}.}}
\vspace*{2mm}

\end{titlepage}

\newpage
\setcounter{page}{2}
\mbox{~}
\newpage

\centerline{\large\bf LHCb collaboration}
\begin{flushleft}
\small
R.~Aaij$^{40}$, 
C.~Abellan~Beteta$^{35,n}$, 
B.~Adeva$^{36}$, 
M.~Adinolfi$^{45}$, 
C.~Adrover$^{6}$, 
A.~Affolder$^{51}$, 
Z.~Ajaltouni$^{5}$, 
J.~Albrecht$^{9}$, 
F.~Alessio$^{37}$, 
M.~Alexander$^{50}$
% ...

{\footnotesize \it
% ...
$ ^{5}$Clermont Universit\'{e}, Universit\'{e} Blaise Pascal, CNRS/IN2P3, LPC, Clermont-Ferrand, France\\
$ ^{6}$CPPM, Aix-Marseille Universit\'{e}, CNRS/IN2P3, Marseille, France\\
% ...
$ ^{9}$Fakult\"{a}t Physik, Technische Universit\"{a}t Dortmund, Dortmund, Germany\\
% ...
$ ^{35}$Universitat de Barcelona, Barcelona, Spain\\
$ ^{36}$Universidad de Santiago de Compostela, Santiago de Compostela, Spain\\
$ ^{37}$European Organization for Nuclear Research (CERN), Geneva, Switzerland\\
% ...
$ ^{40}$Nikhef National Institute for Subatomic Physics, Amsterdam, The Netherlands\\
% ...
$ ^{45}$H.H. Wills Physics Laboratory, University of Bristol, Bristol, United Kingdom\\
% ...
$ ^{50}$School of Physics and Astronomy, University of Glasgow, Glasgow, United Kingdom\\
$ ^{51}$Oliver Lodge Laboratory, University of Liverpool, Liverpool, United Kingdom\\
% ...
$ ^{n}$LIFAELS, La Salle, Universitat Ramon Llull, Barcelona, Spain\\
% ...
}
\end{flushleft}

\cleardoublepage

\renewcommand{\thefootnote}{\arabic{footnote}}
\setcounter{footnote}{0}

\pagestyle{plain}
\setcounter{page}{1}
\pagenumbering{arabic}

\noindent The $\Bcp$ meson is composed of two heavy valence quarks, and has a wide range of expected decay modes\cite{Ivanov:2002un,Ivanov:2006ni,Gouz:2002kk,Kiselev:2000pp,Naimuddin:2012dy,Chang:1992pt,Ebert:2003cn,Ebert:2003wc,AbdElHady:1999xh,Colangelo:1999zn}. Prior to \lhcb taking data, only a few decay channels, such as $\bcp$ and $\Bcp\to\jpsi\mu^+\nu$ had been observed\cite{Abe:1998wi,Abazov:2008kv}. For $pp$ collisions at a centre-of-mass energy of 7\tev, the total \Bcp production cross-section is predicted to be about 0.4\mub, one order of magnitude higher than that at the \tevatron\cite{Chang:2003cr,Gao:2010zzc}. \lhcb has thus been able to observe new decay modes, such as $\Bcp\to\jpsi\pip\pim\pip$\cite{LHCb:2012ag}, $\Bcp\to\psitwos\pip$\cite{Aaij:2013oya} and $\Bcp\to\jpsi D_s^{(*)+}$\cite{Aaij:2013gia}, and to measure precisely the mass of the $\Bcp$ meson\cite{Aaij:2012dd}.

\begin{figure}[b]
 \begin{center}
  \includegraphics[width=0.6\linewidth]{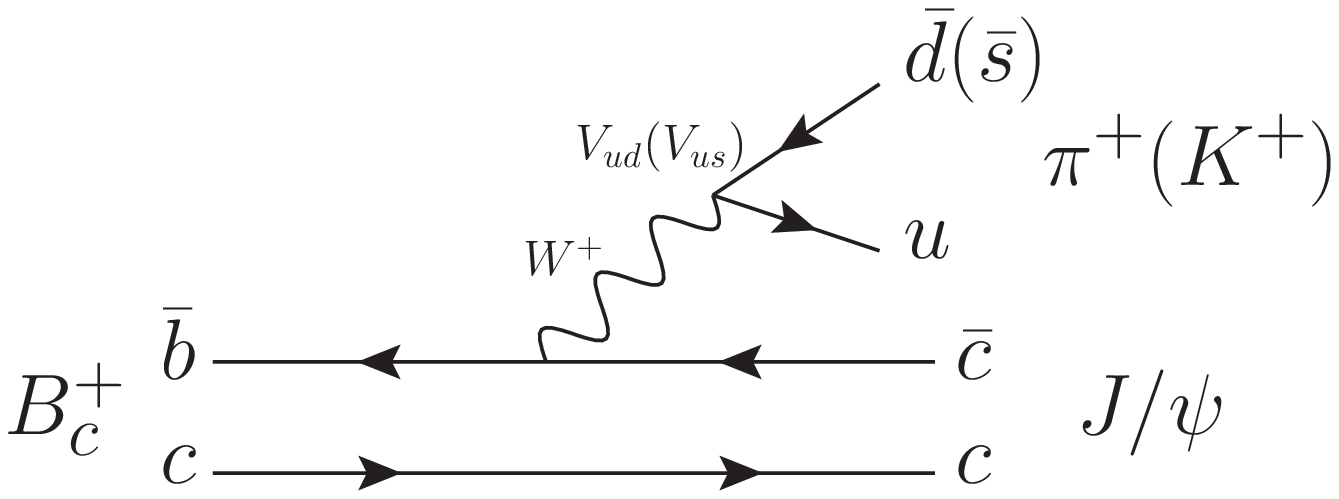}
 \end{center}
 \caption{
 \small
 Diagram for a $\bcp(\Kp)$ decay.
 \label{fig:Bc2JpsiK}
 }
\end{figure}

In this paper, we report the first observation of the decay channel $\bck$ (inclusion of charge conjugate modes is implied throughout the paper). The \jpsi meson is reconstructed in the dimuon final state. The branching fraction is measured relative to that of the $\bcp$ decay mode, which has identical topology and similar kinematic properties, as shown in Fig. \ref{fig:Bc2JpsiK}. No absolute branching fraction of the \Bcp meson is known to date. The predicted ratio of branching fractions $\BR(\bck)/\BR(\bcp)$ is dominated by the ratio of the relevant Cabibbo-Kobayashi-Maskawa (CKM) matrix elements $\left|V_{ud}/V_{us}\right|^2\approx0.05$\cite{Beringer:1900zz}. However, after including the decay constants, $f_{\Kp(\pip)}$, the ratio is enhanced,
\begin{eqnarray}\label{eqn:naiveprediction}
{\BR(\bck)\over\BR(\bcp)}\approx\left|{V_{us}f_{\Kp}\over V_{ud}f_{\pip}}\right|^2=0.077\,,
\end{eqnarray}
where the values of $f_{\Kp(\pip)}$ are given in Ref. \cite{Beringer:1900zz}. Taking into account the contributions of the \Bcp form factor and the kinematics, the theoretical predictions for the ratio of branching fractions lie in the range from 0.054 to 0.088 \cite{Ivanov:2006ni,Gouz:2002kk,Naimuddin:2012dy,Chang:1992pt,Ebert:2003cn,AbdElHady:1999xh,Colangelo:1999zn}. The large span of these predictions is due to the various models and the uncertainties on the phenomenological parameters. The measurement of $\BF(\bck)/\BF(\bcp)$ therefore provides a test of the theoretical predictions of hadronisation.

The analysis is based on a data sample, corresponding to an integrated luminosity of 1.0\invfb of $pp$ collisions, collected by the \lhcb experiment at a centre-of-mass energy  of 7\tev. The \lhcb detector\cite{Alves:2008zz} is a single-arm, forward spectrometer covering the pseudorapidity range $2<\eta<5$ and is designed for precise measurements in the $b$ and $c$ quark sectors. The detector includes a high precision tracking system consisting of a silicon-strip vertex detector surrounding the $pp$ interaction region, a large area silicon-strip detector located upstream of a dipole magnet with a bending power of about 4\tm, and three stations of silicon-strip detectors and straw drift tubes placed downstream. The combined tracking system has momentum resolution $\Delta p/p$ that varies from 0.4\% at 5\gevc to 0.6\% at 100\gevc, and impact parameter (IP) resolution of 20\mum for tracks with high transverse momentum (\pt). Charged hadrons are identified using two ring-imaging Cherenkov (\rich) detectors and good kaon-pion separation is achieved for tracks with momentum between 5\gevc and 100\gevc \cite{LHCb-DP-2012-003}. Photon, electron and hadron candidates are identified by a calorimeter system consisting of scintillating-pad and preshower detectors, an electromagnetic calorimeter and a hadronic calorimeter. Muons are identified by a system composed of alternating layers of iron and multiwire proportional chambers. The trigger system\cite{LHCb-DP-2012-004} consists of a hardware stage, based on information from the calorimeter and muon systems, followed by a two-stage software trigger that applies event reconstruction and reduces the event rate from 1\mhz to around 3\khz.

In the hardware trigger, events are selected by requiring a single muon or dimuon candidate with high \pt. In the software trigger, events are selected by requiring dimuon candidates with invariant mass close to the known $\jpsi$ mass\cite{Beringer:1900zz} and with decay length significance greater than 3 with respect to the associated primary vertex (PV). For events with several PVs, the one with the smallest \chisqip is chosen, where \chisqip is defined as the difference in \chisq of a given PV reconstructed with and without the considered track. The bachelor hadrons ($\Kp$ for $\bck$ and $\pip$ for $\bcp$ decays) are required to be separated from the \Bcp PV and have $\pt>0.5\gevc$. The $\Bcp$ candidates are required to have good vertex quality with vertex fit \chisqvtx per degree of freedom less than 5, and mass within $500\mevcc$ of the world average value of the \Bcp mass \cite{Beringer:1900zz}.

A boosted decision tree (BDT)\cite{Breiman} is used for the final event selection. The BDT is trained using a simulated $\bcp$ sample as a proxy for signal and the high-mass sideband ($m_{\jpsi\pip}>6650\mevcc$) in data for background. The BDT cut value is optimised to maximise the expected $\bck$ signal significance. In the simulation, $pp$ collisions are generated using
\pythia~6.4~\cite{Sjostrand:2006za} with a specific \lhcb
configuration~\cite{LHCb-PROC-2010-056}. The \Bcp meson production is simulated with the dedicated generator \bcvegpy\cite{Chang:2005hq}. Decays of hadronic particles are described by \evtgen~\cite{Lange:2001uf}, in which final state radiation is generated using \photos~\cite{Golonka:2005pn}. The
interaction of the generated particles with the detector and its response are implemented using the \geant toolkit~\cite{Allison:2006ve, *Agostinelli:2002hh} as described in Ref.~\cite{LHCb-PROC-2011-006}. The BDT takes the following variables into account: the \chisqip of the bachelor hadron and \Bcp mesons with respect to the PV; the $\Bcp$ vertex quality; the distance between the $\Bcp$ decay vertex and the PV; the $\pt$ of the $\Bcp$ candidate; the $\chisq$ from the $\Bcp$ decay vertex refit\cite{Hulsbergen:2005pu}, obtained with a constraint on the PV and the reconstructed $\jpsi$ mass; and the cosine of the angle between the momentum of the $\Bcp$ meson and the direction vector from the PV to the \Bcp decay vertex. These variables are chosen as they discriminate the signal from the background, and have similar distributions for $\bck$ and $\bcp$ decays, ensuring that the systematic uncertainty due to the relative selection efficiency is minimal. After the BDT selection, no event with multiple candidates remains.

The branching fraction ratio is computed as
\begin{eqnarray}
{\BF(\bck)\over\BF(\bcp)}={N(\bck)\over N(\bcp)}\cdot{\epsilon(\bcp)\over\epsilon(\bck)}\, ,
\end{eqnarray}
where $N$ is the signal yield of $\bck$ or $\bcp$ decays and $\epsilon$ is the total efficiency, which takes into account the geometrical acceptance, detection, reconstruction, selection and trigger effects.

An unbinned maximum likelihood fit is used to determine the yields from the $\jpsi\Kp$ mass distribution of the \Bcp candidates, under the kaon mass hypothesis. The total probability density function for the fit has four components: signals for $\bck$ and $\bcp$ decays; the combinatorial background; and the partially reconstructed background.

To discriminate between pion and kaon bachelor tracks, the quantity
\begin{eqnarray}
\dllkpi=\ln\mathcal{L}(K)-\ln\mathcal{L}(\pi)
\end{eqnarray}
is used, where $\mathcal{L}(K)$ and $\mathcal{L}(\pi)$ are the
likelihood values provided by the \rich system under the kaon and
pion hypotheses, respectively.
Since the momentum spectra of the
bachelor pions and kaons are correlated with the \dllkpi, the shapes
of the mass distribution used in the fit vary as a function of
\dllkpi.
To reduce this dependence and separate the two signals,
the \dllkpi range is divided into four bins,
$\dllkpi<-5$, $-5<\dllkpi<0$, $0<\dllkpi<5$ and $\dllkpi>5$.
The ratio of the total signal yields is defined as
$\mathcal{R}_{\Kp/\pip}=\sum^{4}_{i=1} N^{i}_{\jpsi\Kp}/ \sum^{4}_{i=1} N^{i}_{\jpsi\pip}$,
where
$N_{\jpsi\Kp(\pip)}^{i}$ is the signal
yield in each \dllkpi bin $i$.
Due to the limited sample size of the
$\bck$ signal in the bins with $\dllkpi<-5$ and $-5<\dllkpi<0$,
their signal yields are fixed, respectively, to be zero and $P\times \sum^{4}_{i=1} N^{i}_{\jpsi\Kp}$
where the $P$ is the probability that the kaon from the $\bck$ decay has
$-5<\dllkpi<0$, as estimated from simulation.

\begin{figure}[t]

 \begin{center}

  \includegraphics[width=0.50\linewidth]{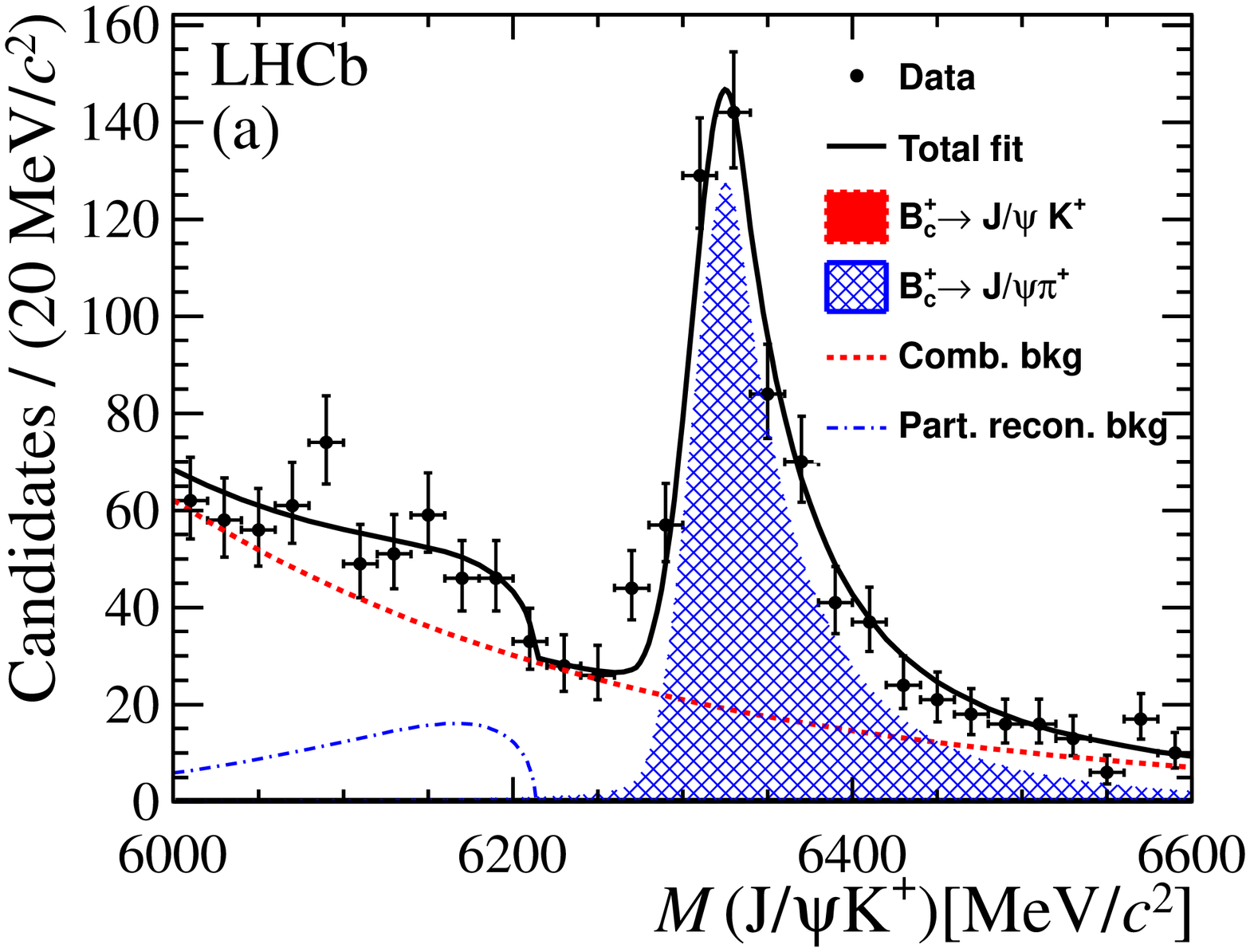}%
  \hspace{-0.06cm}%
  \includegraphics[width=0.50\linewidth]{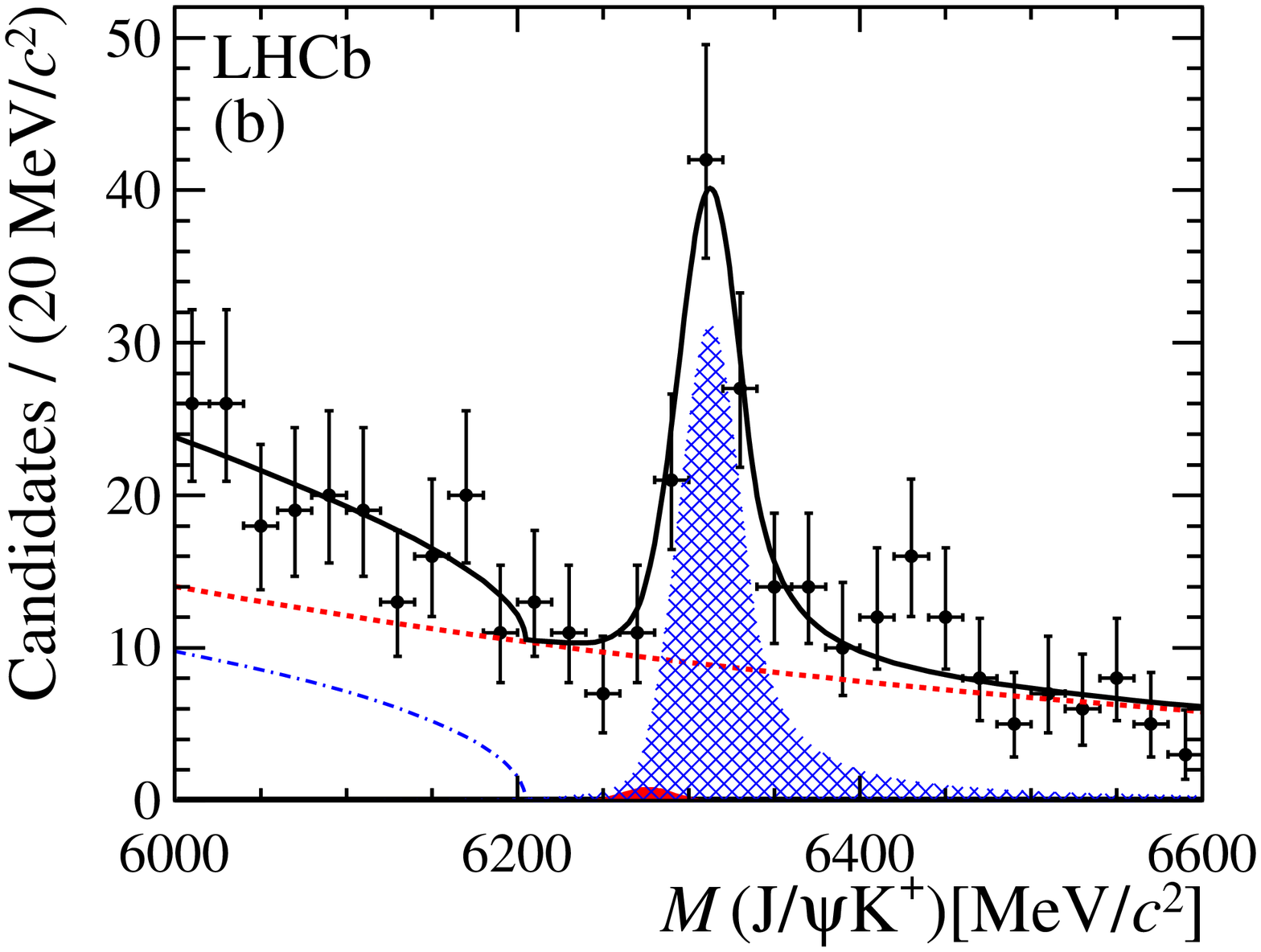}\\
  \vspace{-0.1cm}
  \includegraphics[width=0.50\linewidth]{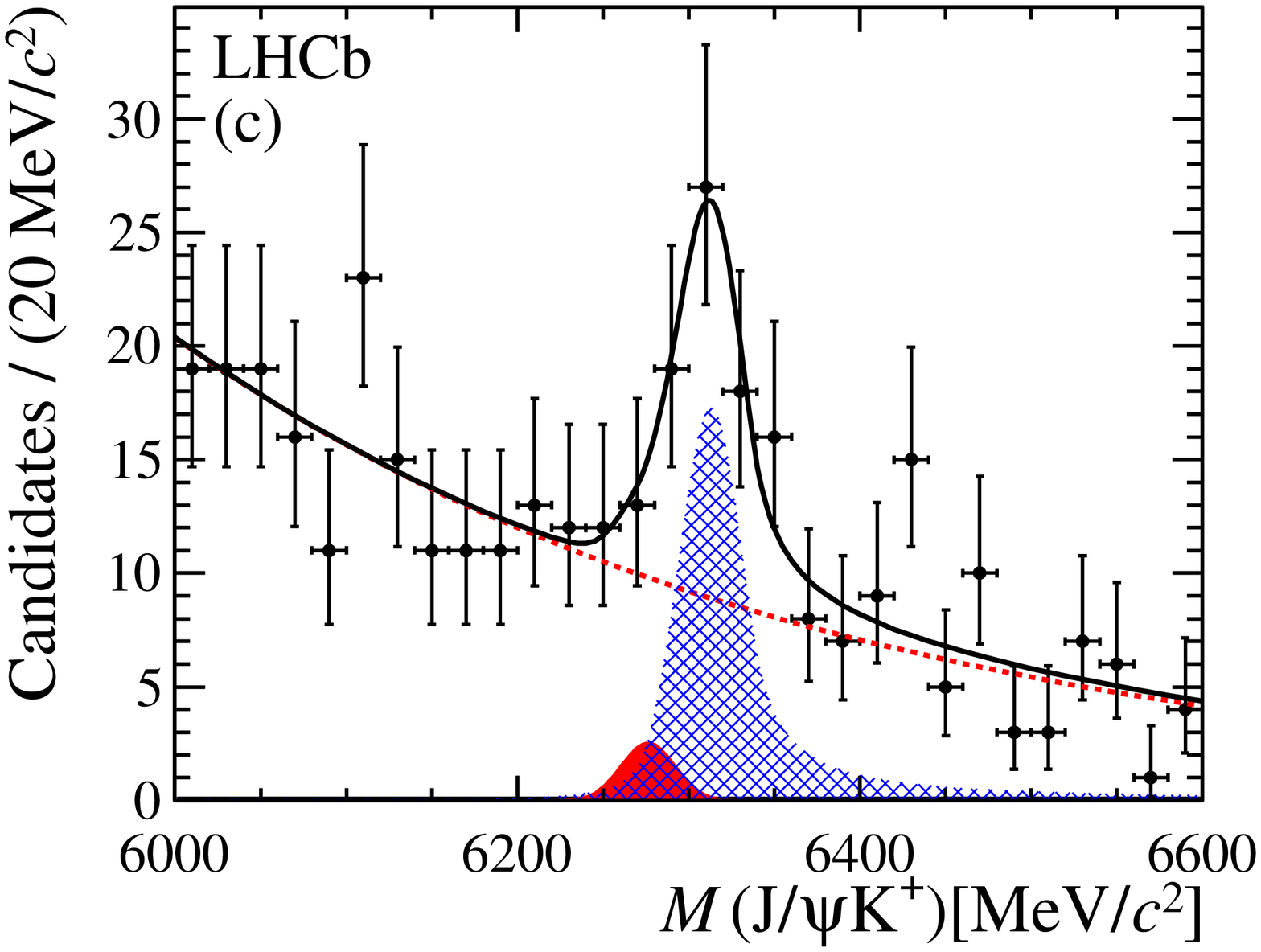}%
  \hspace{-0.06cm}%
  \includegraphics[width=0.50\linewidth]{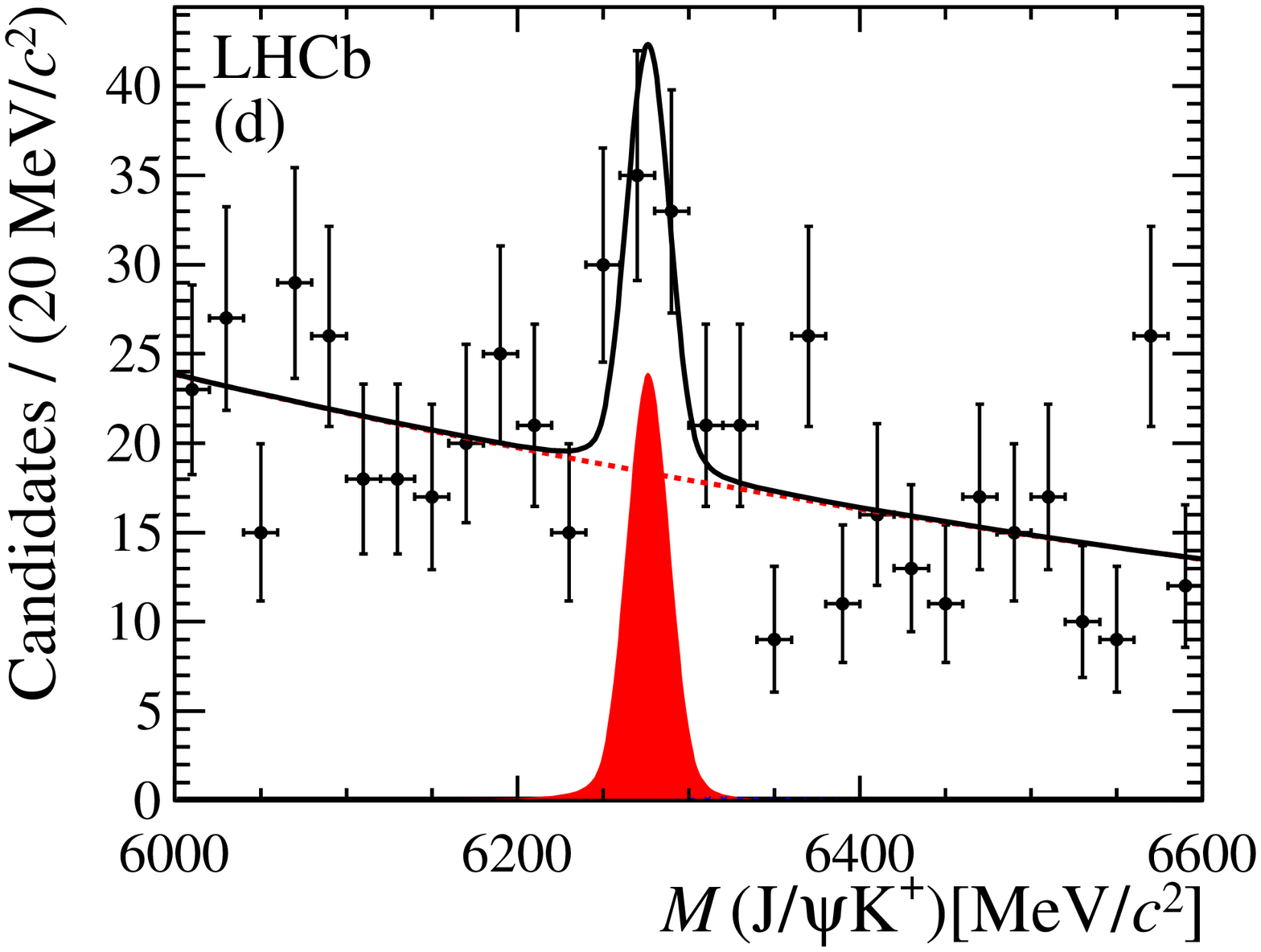}\\
  \vspace*{-0.25cm}
 \end{center}

 \caption{
 \small
 Mass distributions of \Bcp candidates in four \dllkpi bins and the superimposed fit results. The solid shaded area (red) represents the $\bck$ signal and the hatched area (blue) the $\bcp$ signal. The dot-dashed line (blue) indicates the partially reconstructed background and the dotted (red) the combinatorial background. The solid line (black) represents the sum of the above components and the points with error bars (black) show the data. The labels $(a)$, $(b)$, $(c)$ and $(d)$ correspond to $\dllkpi<-5$, $-5<\dllkpi<0$, $0<\dllkpi<5$ and $\dllkpi>5$ for the bachelor track, respectively.
 \label{fig:Bc2JpsiK_Fit}
 }
\end{figure}

Figure \ref{fig:Bc2JpsiK_Fit} shows the invariant mass distributions of the \Bcp candidates, calculated with the kaon mass hypothesis in the four \dllkpi bins. In the fit to the \Bcp mass spectrum, the shape of the $\bck$ signal is modelled by a double-sided Crystal Ball (DSCB) function\cite{Skwarnicki:1986xj} as
\begin{eqnarray}
 f(x;M,\sigma,a_l,n_l,a_r,n_r)=
 \left\{
 \begin{aligned}
  & e^{-a^2_l\over2}\left({n_l\over a_l}\right)^{n_l}\left({n_l\over a_l}-a_l-{x-M\over\sigma}\right)^{-n_l} & {x-M\over\sigma}<-a_l\\
  & \exp\left[-{1\over2}\left({x-M\over\sigma}\right)^2\right] & -a_l\leq{x-M\over\sigma}\leq-a_r\\
  & e^{-a_r^2\over2}\left({n_r\over a_r}\right)^{n_r}\left({n_r\over a_r}-a_r+{x-M\over\sigma}\right)^{-n_r} & {x-M\over\sigma}>-a_r\\
 \end{aligned}
 \right.
\end{eqnarray}
where the peak position is fixed to that from an independent fit to the $\bcp$ mass distribution, and the tail parameters $a_{l,r}$ and $n_{l,r}$ on both sides are taken from simulation.

As the decay $\bcp$ is reconstructed with the kaon mass replacing the pion mass, the signal is shifted to higher mass values and is modelled by another DSCB function whose shape and the relative position to the $\bck$ signal are also derived from simulation. Two corrections are applied to the $\bcp$ simulation sample. Firstly, since the resolution of the detector is overestimated, the momenta of charged particles are smeared to make the resolution on the \Bcp mass in the $\bcp$ simulation sample the same as that of the $\jpsi\pip$ mass distribution of the \Bcp candidates in the data sample. Secondly, the shapes of the $\bcp$ mass distribution in the four \dllkpi bins depend on the \dllkpi distribution, which is different in data and simulation. To reduce the effect of this difference, each simulated event is reweighted by a \dllkpi dependent correction factor, which is derived from a linear fit to the ratio of the \dllkpi distribution in background-subtracted data, to that of the simulation sample. The background subtraction \cite{Pivk:2004ty} is performed with the $\jpsi\pip$ mass distribution of the \Bcp candidates in the data sample with the pion mass hypothesis.

The combinatorial background is modelled as an exponential function with a different freely varying parameter in each \dllkpi bin. The contribution of the partially reconstructed background is modelled by an ARGUS function\cite{Albrecht:1990am}. The contribution of the partially reconstructed background is dominated by events with bachelor pions, which are suppressed in the high-value \dllkpi bins, therefore the number of the partially reconstructed events in the $\dllkpi>5$ bin is assumed to be zero. All parameters of the partially reconstructed background are allowed to vary. The observed $\bck$ signal yield is $46\pm12$ and the ratio of yields is
\begin{eqnarray*}
\mathcal{R}_{\Kp/\pip}={N(\bck)\over N(\bcp)}=0.071\pm0.020\,\stat\,.
\end{eqnarray*}

The ratio of the total efficiencies computed over the full \dllkpi range is
\begin{eqnarray*}\label{eqn:efficiency}
{\epsilon(\bck)\over\epsilon(\bcp)}=1.029\pm0.007\, ,
\end{eqnarray*}
which is determined from simulation and the uncertainty is due to the finite size of the simulation samples.

The $\bcp$ signal has a long tail that may extend into the high mass region. A systematic uncertainty is assigned due to the choice of fit range, and is determined to be 0.9\% by changing the mass window from 6000-6600 \mevcc to 6200-6700\mevcc and comparing the results. To estimate the systematic uncertainty due to the potentially different performance of the BDT on data and simulation, several BDT cut values were tested. A 5.7\% spread in the final result is obtained and is propagated to the quoted systematic uncertainty.

To estimate the uncertainty due to the shapes of the $\bck$ and $\bcp$ signals, the fit is repeated many times by varying the parameters of the tails of these DSCB functions that were kept constant in the fit within one standard deviation of their values in simulation. A spread of 0.7\% is observed. For the $\bcp$ signal the assigned systematic uncertainty is 0.5\%.

To estimate the systematic uncertainty due to the choice of signal shape, an alternative $\bcp$ mass shape is used, which is determined from the data sample by subtracting the background in the $\jpsi\pip$ mass distribution of the \Bcp candidates with the pion hypothesis. A $2.7\%$ difference with the ratio obtained with the nominal signal shape is observed.

For the systematic uncertainty due to the choice of the partially reconstructed background shape in each \dllkpi bin, the shape is modelled with the ARGUS function convolved with a Gaussian function. The observed 2.3\% deviation from the default fit is assigned as the systematic uncertainty.

For the $\bck$ yields in the two bins with $\dllkpi<0$, half of the probability estimated from the simulation, namely 1.8\%, is taken as systematic uncertainty.

To estimate the uncertainty due to the choice of the \dllkpi binning, two other binning choices are tried: $\dllkpi<-6$, $-6<\dllkpi<-1$, $-1<\dllkpi<4$, $\dllkpi>4$ and $\dllkpi<-4$, $-4<\dllkpi<1$, $1<\dllkpi<6$, $\dllkpi>6$. The average value of the results with these two binning choices has a 1.2\% deviation from the default value, which is taken as the systematic uncertainty.

There is a systematic uncertainty due to the different track reconstruction efficiencies for kaons and pions. Since the simulation does not describe hadronic interactions with detector material perfectly, a 2\% uncertainty is assumed, as in Ref.\cite{Aaij:2012jw}.

An uncertainty of 0.7\% arises from the statistical uncertainty of the ratio of the total efficiencies, which is due to the finite size of the simulation sample.

\begin{table}[t]\label{tab:systematic}
 \caption{\small Relative systematic uncertainties on the ratio of branching fractions.}
 \begin{center}
 \begin{tabular}{lc}
  Source & Uncertainty (\%) \\
  \hline
  Mass window & 0.9 \\
  BDT selection & 5.7 \\
  $\bck$ signal model & 0.7 \\
  $\bcp$ signal model & 0.5 \\
  Choice of signal shape & 2.7 \\
  Partially reconstructed background shape & 2.3 \\
  $\bck$ signals in $\dllkpi<0$ bins & 1.8 \\
  $\dllkpi$ binning choice & 1.2 \\
  $\Kp$ and $\pip$ interaction length & 2.0 \\
  Simulation sample size & 0.7 \\
  \hline
  Total & 7.5\\
 \end{tabular}
 \end{center}
\end{table}

The systematic uncertainties are summarised in Table \ref{tab:systematic}. The total systematic uncertainty, obtained as the quadratic sum of the individual uncertainties, is 7.5\%.

The asymptotic formula for a likelihood-based test $\sqrt{-2\ln(\mathcal{L}_\mathrm{B}/\mathcal{L}_\mathrm{S+B})}$ is used to estimate the $\bck$ signal significance, where $\mathcal{L}_\mathrm{B}$ and $\mathcal{L}_\mathrm{S+B}$ stand for the likelihood of the background-only hypothesis and the signal and background hypothesis respectively. A deviation from the background-only hypothesis with 5.2 standard deviations is found when only the statistical uncertainty is considered. When taking the systematic uncertainty into account, the total significance of the $\bck$ signal is 5.0 $\sigma$.

In summary, a search for the $\bck$ decay is performed using a data sample, corresponding to an integrated luminosity of 1.0 \invfb of $pp$ collisions, collected by the \lhcb experiment. The signal yield is $46\pm12$ candidates, and represents the first observation of this decay channel. The branching fraction of $\bck$ with respect to that of $\bcp$ is measured as
\begin{eqnarray*}\label{eqn:finalresult}
{\BF(\bck)\over\BF(\bcp)}=0.069\pm0.019\pm0.005\, ,
\end{eqnarray*}
where the first uncertainty is the statistical and the second is systematic. The measurement is in agreement with the theoretical predictions\cite{Ivanov:2006ni,Gouz:2002kk,Naimuddin:2012dy,Chang:1992pt,Ebert:2003cn,AbdElHady:1999xh,Colangelo:1999zn}.

Assuming factorization holds, the na\"ive prediction of the ratio $\BR(\bck)/\BR(\bcp)$ can be compared to other \B meson decays with a similar topology
\begin{eqnarray}\label{eqn:branchingratio}
 {\BR(\B\to\D\Kp)\over\BR(\B\to\D\pip)}=\left\{
 \begin{aligned}
  &0.0646\pm0.0043\pm0.0025 & \mathrm{for}\, & \Bs\to\Dsm\Kp(\pip)\\
  &0.0774\pm0.0012\pm0.0019 & \mathrm{for}\, & \Bu\to\Dzb\Kp(\pip)\\
  &0.074\pm0.009 & \mathrm{for}\, & \Bd\to\Dm\Kp(\pip)\\
 \end{aligned}
 \right.
\end{eqnarray}
taken from Ref.\cite{Aaij:2012zz,Aaij:2012kz,Beringer:1900zz}. Hence, this measurement of $\BR(\bck)/\BR(\bcp)$ is consistent with na\"ive factorisation in \B decays.

\section*{Acknowledgements}

\noindent We express our gratitude to our colleagues in the CERN accelerator departments for the excellent performance of the LHC. We thank the technical and administrative staff at the LHCb institutes. We acknowledge support from CERN and from the national agencies: CAPES, CNPq, FAPERJ and FINEP (Brazil); NSFC (China); CNRS/IN2P3 and Region Auvergne (France); BMBF, DFG, HGF and MPG (Germany); SFI (Ireland); INFN (Italy); FOM and NWO (The Netherlands); SCSR (Poland); MEN/IFA (Romania); MinES, Rosatom, RFBR and NRC ``Kurchatov Institute'' (Russia); MinECo, XuntaGal and GENCAT (Spain); SNSF and SER (Switzerland); NAS Ukraine (Ukraine); STFC (United Kingdom); NSF (USA). We also acknowledge the support received from the ERC under FP7. The Tier1 computing centres are supported by IN2P3 (France), KIT and BMBF (Germany), INFN (Italy), NWO and SURF (The Netherlands), PIC (Spain), GridPP (United Kingdom). We are thankful for the computing resources put at our disposal by Yandex LLC (Russia), as well as to the communities behind the multiple open
 source software packages that we depend on.

\addcontentsline{toc}{section}{References}
\bibliographystyle{LHCb}
\bibliography{main}

\end{document}